# Alloyed Re$_x$Mo$_{1-x}$S$_2$ Nanoflakes with Enlarged Interlayer Distances for Hydrogen Evolution

Jing Li,[1] René Hübner,[2] Marielle Deconinck,[3,4] AnkitaBora,[1] Markus Göbel,[5] Dana Schwarz,[6] Guangbo Chen,[7] Guangzhao Wang,[8,*] Shengyuan A. Yang,[9] Yana Vaynzof,[3,4] Vladimir Lesnyak[1,*]

[1] Physical Chemistry, TU Dresden, Zellescher Weg 19, 01069 Dresden, Germany
[2] Institute of Ion Beam Physics and Materials Research, Helmholtz-Zentrum Dresden-Rossendorf e.V., Bautzner Landstrasse 400, 01328 Dresden, Germany
[3] Chair for Emerging Electronic Technologies, TU Dresden, Nöthnitzer Str. 61, 01187 Dresden, Germany
[4] Leibniz-Institute for Solid State and Materials Research Dresden, Helmholtzstraße 20, 01069 Dresden, Germany
[5] Electrochemistry, TU Dresden, Zellescher Weg 19, 01069 Dresden, Germany
[6] Leibniz-Institut für Polymerforschung Dresden e.V., Hohe Straße 6, 01069 Dresden, Germany
[7] Center for Advancing Electronics Dresden (cfaed) and Faculty of Chemistry and Food Chemistry, TU Dresden, 01062 Dresden, Germany
[8] School of Electronic Information Engineering, Yangtze Normal University, 16 Juxian Road, Fuling district, 408100 Chongqing, China
[9] Research Laboratory for Quantum Materials, Singapore University of Technology and Design, 8 Somapah Road, Singapore 487372, Singapore

Molybdenum sulfide (MoS$_2$) has attracted significant attention due to its great potential as a low-cost and efficient catalyst for the hydrogen evolution reaction. Developing a facile, easily upscalable, and inexpensive approach to produce catalytically active nanostructured MoS$_2$ with a high yield would significantly advance its practical application. Colloidal synthesis offers several advantages over other preparation techniques to overcome the low reaction yield of exfoliation and drawbacks of expensive equipment and processes used in chemical vapor deposition. In this work, we report an efficient synthesis of alloyed Re$_x$Mo$_{1-x}$S$_2$ nanoflakes with an enlarged interlayer distance, among which the composition Re$_{0.55}$Mo$_{0.45}$S$_2$ exhibits excellent catalytic performance with overpotentials as low as 79 mV at 10 mA/cm$^2$ and a small Tafel slope of 42 mV/dec. Density functional theory calculations prove that enlarging the distance between layers in the Re$_x$Mo$_{1-x}$S$_2$ alloy can greatly improve its catalytic performance due to a significantly reduced free energy of hydrogen adsorption. The developed approach paves the way to design advanced transition metal dichalcogenide-based catalysts for hydrogen evolution and to promote their large-scale practical application.

**Introduction**

Hydrogen ($H_2$) is considered as one of the most potent energy forms due to its highest mass energy density, storability, and renewability. (1−4) Electro-catalytic reduction of water through the hydrogen evolution reaction (HER) is one of the most efficient and sustainable strategies for $H_2$ generation. (5−7) In this process, the adsorption energy of hydrogen is one of the most important conditions for selecting a catalyst. Presently, platinum is the best-known catalyst due to its close-to-zero free energy for adsorbing hydrogen atoms, but its high cost limits its extensive practical application. (8,9) Therefore, we need to develop a catalyst made of abundant and relatively cheap materials. (10−12) Among potential candidates, transition metal dichalcogenides (TMDs) have extensively been investigated for the electro-catalytic production of hydrogen, where the efficiency of hydrogen generation has remarkably been promoted, leading in turn to a significant decrease of the price for the catalyst while achieving its high performance. (1,5,6,13−15)

Typically, TMDs with the formula of $MX_2$ (M = metal from groups IV–VIB and X = S, Se, or Te) have a layered structure, where X–M–X layers are bound by weak van der Waals forces. (16) Among presently developed TMDs, $MoS_2$ is the most studied compound due to its stability in acidic environments and low price. (17−19) Its HER activity is, however, limited because only the edge sites of the $MoS_2$ layers are catalytically active owing to their appropriate free energy of hydrogen adsorption ($\Delta G_H$ = −0.45 eV), while the basal planes remain catalytically inert ($\Delta G_H$ = 1.92 eV). (6,20−22) Therefore, extensive research efforts have been devoted to improve the electro-catalytic performance of $MoS_2$. (5,21−25) For example, recently, it was demonstrated that control of the interlayer distance in $MoS_2$ is an efficient way to improve its electrochemical performance. (8,11,26−28) Changing the interlayer distance in $MoS_2$ nanosheets can modulate the local chemical environment and electronic structure, resulting in an enhancement of the catalytic activity on the edge sites. Thus, the Xie group found that a lower synthesis temperature could lead to the formation of Mo–O bonds inherited from the molybdate precursor, resulting in oxygen-incorporated $MoS_2$ nanosheets with enlarged interlayer distance, which exhibited an improved catalytic performance. (7) The Sun group also found that due to the intercalation of oxidized dimethylformamide (DMF) species between two S–Mo–S layers, the interlayer distance was enlarged, and $MoS_2$ with larger interlayer distance led to a better hydrogen absorption and could provide a denser H coverage on the edge sites than non-expanded $MoS_2$, which, in turn, resulted in a better HER performance. (8) Furthermore, the larger interlayer distance induced ion transport resistance in the layer space, which also elevated the HER performance. These examples demonstrate that enlarging the interlayer distance in MoS$_2$ can significantly improve its catalytic properties.

The incorporation of atoms of the transition metal rhenium (Re) into MoS$_2$ is another common strategy to increase the number of catalytically active sites to improve the HER performance. Yang et al. found that it is possible to make also basal planes of MoS$_2$ catalytically active by incorporation of Re atoms to form alloyed Re$_{0.55}$Mo$_{0.45}$S$_2$ with a distorted 1T structure. (29) Density functional theory (DFT) calculations revealed the change in Gibbs free energy associated with hydrogen adsorption from 2H MoS$_2$ to distorted 1T MoS$_2$, resulting in improvement of the HER performance. Kochat et al. demonstrated that S atoms in the alloyed Re$_x$Mo$_{1-x}$S$_2$ phase become more activated as compared to S atoms in the pure MoS$_2$ phase, and a sample with 55% Re exhibited the best HER performance among other alloy compositions. (30) The Kang group attributed an enhanced catalytic activity of the Re$_{0.5}$Mo$_{0.5}$S$_2$ alloy to the formation of S–H or Mo–H bonds at one of the S vancancies. (31) Combining these two strategies, i.e., the interlayer distance enlargement and the alloy formation, may further promote the catalytic activity of the material.

There are several methods to prepare monolayer or few-layer thick TMD nanomaterials to study their electrocatalytic performance, such as chemical vapor deposition, chemical or physical exfoliation, and colloidal synthesis. (32) Compared to the other synthesis approaches, the colloidal method can overcome the low reaction yield of exfoliation and drawbacks of expensive equipment used in chemical vapor deposition, which make colloidal synthesis a facile and scalable route to TMD-based catalysts. (32) Commonly, colloidal TMD nanoparticles contain long-chain organic ligands, the presence of which can hamper electron transport within the material, thus deteriorating its catalytic performance. (33) Such nanomaterials require additional treatment to remove the ligands and activate the surface. In order to skip this treatment step, we recently developed a general colloidal synthesis of ligand-free TMD nanomaterials by using a polar organic solvent, formamide (FA). (34) We believe that this new synthesis method has great potential in simplifying the procedure and, at the same time, improving the catalytic performance of the resulting materials further, bringing it closer to industrial application.

In this work, we employ the new method and demonstrate the synthesis of colloidal-alloyed Re$_x$Mo$_{1-x}$S$_2$ nanomaterials with enlarged interlayer distances, which simultaneously implement both strategies to promote catalytic activity. Polar formamide used as a solvent can form donor–acceptor complexes with metal precursors, which contribute to the enlarged interlayer distance in growing thin nanoparticles. At the same time, its presence on the surface does not block active sites available for the efficient catalysis. We tested different sulfur precursors aiming at the formation of small particles. Among them, $MoS_2$ nanoparticles synthesized by using alkali metal sulfides possessed the desired small average lateral size of approx. 10 nm and a thickness of approx. 3 nm. The alloyed sample with the composition $Re_{0.55}Mo_{0.45}S_2$ demonstrated an enhanced HER performance characterized by overpotentials as low as 79 mV at the current density of 10 mA/cm$^2$ and Tafel slopes as small as 42 mV/dec. DFT calculations proved that the Gibbs free energy of hydrogen adsorption on this nanomaterial is remarkably decreased. In addition, the shift of its density of states demonstrates the possibility of tuning the electronic structure of the material. These findings may pave the way for the controllable growth of small nanoparticles made of other TMD compounds with enlarged interlayer distances, which are ready for application in electrocatalysis without an additional treatment.

**Results and Discussion**

**Colloidal Synthesis of TMD Nanomaterials with Enlarged Interlayer Distances**

A schematic overview of the synthesis procedure is shown in Figure 1, and the details are described in the Experimental Section. We adopted a recently developed strategy to grow small and thin TMD nanoparticles in a polar organic solvent without using long-chain organic ligand molecules. (34) As solvents, we tested formamide, *N*-methylformamide, and dimethylformamide (their structures are shown in Figure S1 in the Supporting Information). Each solvent can dissolve two types of precursors, i.e., $MoCl_5$ (or $ReCl_5$) and the S-precursors, forming clear solutions. Here, transition metal ions ($Mo^{5+}$ and $Re^{5+}$) can easily form a donor–acceptor complex with formamide and its derivatives. This type of complexation can also lead to an increase of the interlayer distance in the growing TMD nanoparticles. Since formamide is the least volatile among the three solvents, it has a higher boiling point of 210 °C; we used it in all of the following experiments. We further simplified the original method by replacing the degassing of the precursor solutions with nitrogen bubbling, which does not require a vacuum system and is thus more adaptable for upscaling and industrial application. We first synthesized the pure-phase $MoS_2$ and $ReS_2$ nanomaterials at a moderate temperature of 170 °C, which was expected to deliver small-sized particles with a large contribution of edge site regions to the total surface area, desirable for their catalytic application.

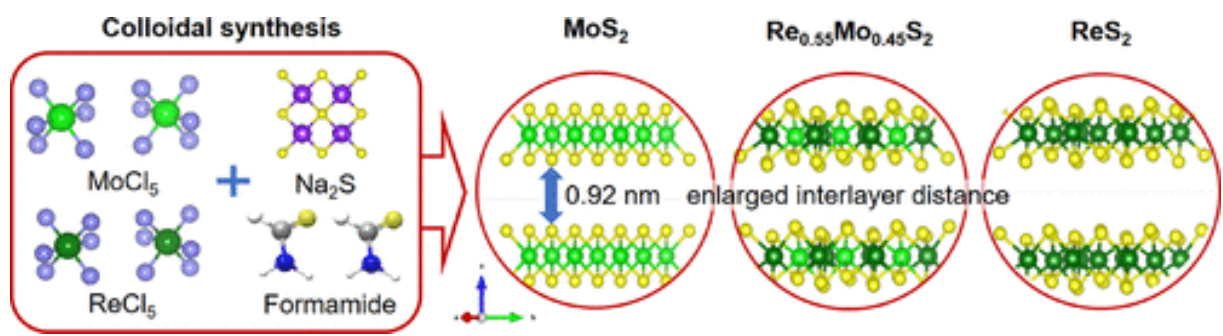

**Figure 1.** Scheme of the synthesis of TMD nanomaterials with enlarged interlayer distances.

In the synthesis of the MoS$_2$ nanoparticles, we explored different S-precursors: thiourea, Li$_2$S, Na$_2$S, and K$_2$S. In all cases, the S-precursors were added in excess to MoCl$_5$ in order to ensure a complete metal conversion and a stoichiometric composition of the material. We observed that when thiourea was used, the color of the reaction mixture changed from orange to black after injecting the Mo-precursor in approx. 3 min, while by using Li$_2$S, Na$_2$S, or K$_2$S, the color changed faster, implying faster reaction kinetics in the case of the alkali metal sulfides. The synthesized MoS$_2$ exhibited a thin sheet-like morphology with lateral sizes of up to tens of nanometers, broad size distributions, and rather irregular shape, as shown in Figure S2. Therefore, we use the term nanoflakes, which better reflects this type of morphology. The flakes tended to crumple together, forming large aggregates in which individual particles were randomly oriented. The thickness of those flakes standing on the grid can be estimated to be 3–5 nm. In the transmission electron microscopy (TEM) images, one can see that by using thiourea, the flakes grew larger than in the case of alkali metal sulfides. Apparently, the salts are more reactive toward the Mo-precursor than thiourea as they readily provide S$^{2-}$ ions being dissolved in FA, while thiourea can release reactive sulfur species only upon breaking covalent bonds, which require additional time and energy. This results in a higher concentration of sulfur species in the reaction mixture in the case of the alkali metal sulfides, which in turn leads to a larger number of nuclei formed and thus to a smaller size of the final particles. Small-sized nanomaterials can provide more active sites for HER, making alkali metal sulfides more suitable for the synthesis. Among the three sulfides tested, Na$_2$S is more stable and cheaper, compared to K$_2$S and Li$_2$S. TEM element mapping of MoS$_2$ nanoflakes revealed a uniform distribution of Mo and S over the particles, as shown in Figures S3–S5. Interestingly, the nanoflakes synthesized with Na$_2$S and K$_2$S contain significant amounts of the corresponding alkali metal, suggesting that it can be present on the MoS$_2$ surface as well as being embedded in between the S–Mo–S monolayers.

Na$_2$S was further utilized in the synthesis of ReS$_2$, resulting in a similar morphology of the product, as shown in Figure S6. As mentioned above, the incorporation of Re atoms into the MoS$_2$ phase can further increase the number of active sites on its surface to improve the HER performance. Therefore, in the next step, we synthesized Re$_x$Mo$_{1-x}$S$_2$ alloy nanoflakes with different compositions by mixing the Re- and Mo-precursors at different ratios, using the reaction conditions optimized in the synthesis of the pure MoS$_2$ samples. As in the case of pure MoS$_2$, alloyed Re$_{0.55}$Mo$_{0.45}$S$_2$ nanoflakes easily agglomerate due to the absence of ligands and their small size, as seen in Figure 2a. From individual particles visible in the TEM images, we can estimate their lateral sizes to be around 5 nm, the thickness to be about 3–5 nm (see Figure 2b), and the layer-to-layer distance to be 0.9 nm (see the inset of Figure 2b). From the high-angle annular dark-field scanning TEM (HAADF-STEM) image in Figure 2c and the corresponding element distribution maps of Mo, Re, and S presented in Figure 2d, one can see a uniform distribution of the three elements over the Re$_{0.55}$Mo$_{0.45}$S$_2$ nanoflakes, confirming their alloy structure.

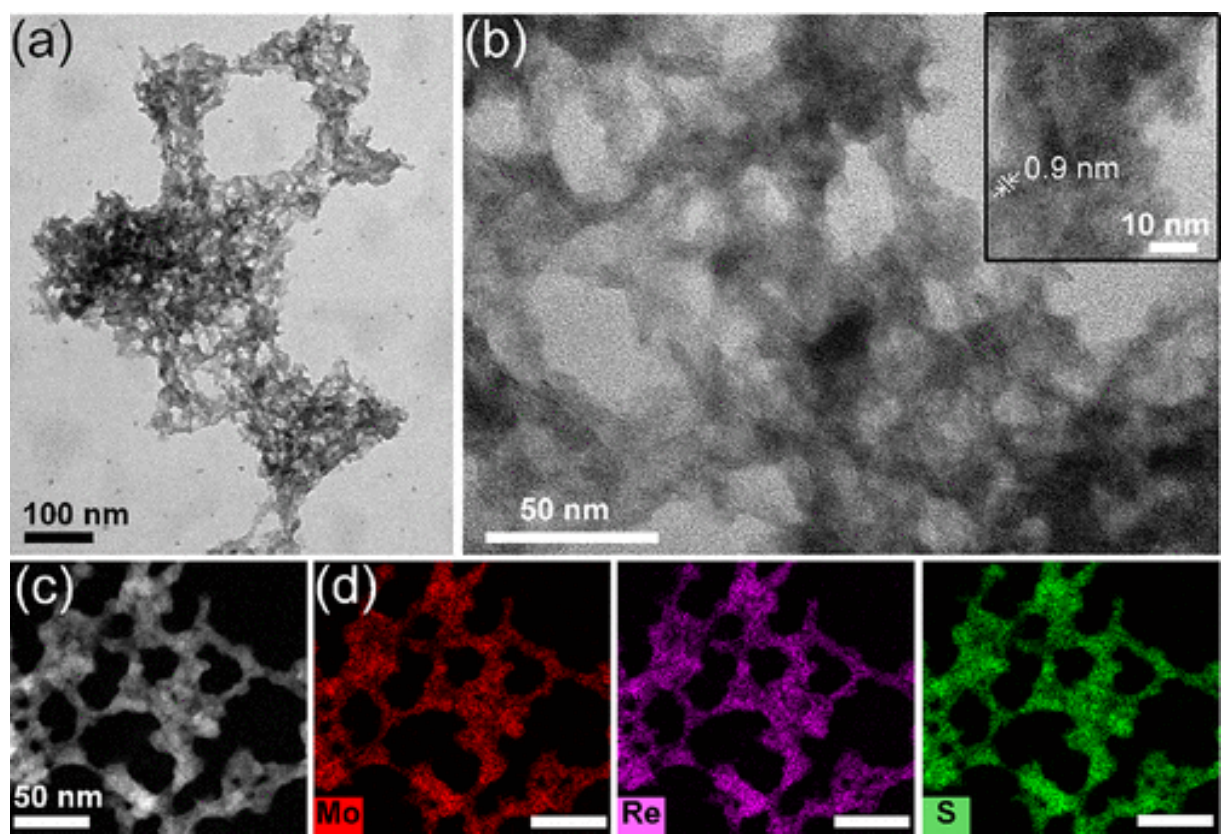

**Figure 2.** Conventional TEM images of Re$_{0.55}$Mo$_{0.45}$S$_2$ nanoflakes at different magnifications (a, b). The inset in (b) shows the interlayer distance. HAADF-STEM image (c) and corresponding EDXS-based element distribution maps (d) of Mo, Re, and S.

Small nanoparticles obtained by the colloidal synthesis at relatively moderate temperature are typically poorly crystalline and contain many surface defects, which can provide more active sites for the HER. However, such small particles full of defects processed in the form of thin films can also lead to decreased electrical conductivity. In order to improve it, we used carbon black as an additive in the synthesis of alloyed $Re_{0.55}Mo_{0.45}S_2$ nanoflakes, aiming at a final weight nanoflakes:carbon ratio of 10:3. In this way, nucleation of the flakes can occur (at least partially) directly on the existing carbon particles acting as templates, providing good contact between $Re_{0.55}Mo_{0.45}S_2$ and the conducting carbon scaffold. Via TEM imaging of the product, we observed rounded hexagonal carbon black particles, as shown in [Figure S7a]. From the HAADF-STEM image in [Figure S7b] and the corresponding element distribution maps of Mo, Re, and S presented in [Figure S7c], one can see that the carbon particles are covered by $Re_{0.55}Mo_{0.45}S_2$ nanoflakes, with a higher concentration at the edge of the carbon disks. To further study the morphology of the samples, we performed scanning electron microscopy (SEM) imaging of drop-cast films of pure $Re_{0.55}Mo_{0.45}S_2$ nanoflakes, $Re_{0.55}Mo_{0.45}S_2$ nanoflakes mixed with carbon black, and the $Re_{0.55}Mo_{0.45}S_2$ nanoflakes/carbon composite (see [Figure S8]). Apparently, the film made of pure $Re_{0.55}Mo_{0.45}S_2$ nanoflakes exhibited the smoothest surface due to their small size and efficient dispersion. With the introduction of carbon black, the $Re_{0.55}Mo_{0.45}S_2$nanoflakes/carbon composite film shows a smoother surface than the nanoflakes mixed with the carbon black sample. In all cases, the films have rather dense microstructures with a small number of holes.

We further investigated the structures and compositions of the products by Raman spectroscopy, inductively coupled plasma optical emission spectroscopy (ICP-OES), and powder X-ray diffraction (XRD) analysis. The Raman spectrum of the $MoS_2$ nanoflakes presented in [Figure S9] shows two distinct peaks at 372 cm$^{-1}$ ($E_{2g}$ mode, attributed to in-plane vibration) and 400 cm$^{-1}$ ($A_{1g}$ mode, attributed to out-of-plane vibration), where the $A_{1g}$ mode experiences a shift to larger wavenumbers as compared to published data. (35,36) This shift indicates a weakened interlayer mechanical coupling strength, suggesting that the interlayer distance is enlarged. (8) Raman spectra of the $ReS_2$ and $Re_{0.55}Mo_{0.45}S_2$ nanoflakes do not exhibit well-resolved peaks (spectra are not shown) probably because of their poorer crystallinity and smaller size. Element analysis via ICP-OES revealed compositions close to stoichiometric in all three types of samples, i.e., in $MoS_2$, the Mo:S ratio was found to be 1:2.03; in $ReS_2$, the Re:S ratio was 1:1.93; and in the alloyed nanoflakes, the (Re + Mo):S ratio was 1:1.96. The crystalline structures of the $MoS_2$, $ReS_2$, and $Re_{0.55}Mo_{0.45}S_2$ nanomaterials were characterized by XRD analysis. As seen in [Figure 3], the XRD patterns of $MoS_2$ and $ReS_2$ are significantly different from the corresponding bulk 2H MoS$_2$ and 1T′ ReS$_2$ references. One main peak appears in the low-angle region at 9.6°, corresponding to the 002 reflection of the nanoflakes. Using Bragg's law $2d\sin\theta = n\lambda$, we can estimate the interlayer distance increase from 0.635 to 0.922 nm. It is worth noting that the 002 peak is quite symmetric and much more pronounced than other reflections from lattice planes not laying in the lateral directions. The peak position for the three different samples is almost the same because it is dictated mainly by the similar interlayer distance. A sharper peak in the case of Re$_{0.55}$Mo$_{0.45}$S$_2$ suggests a larger crystallite size.

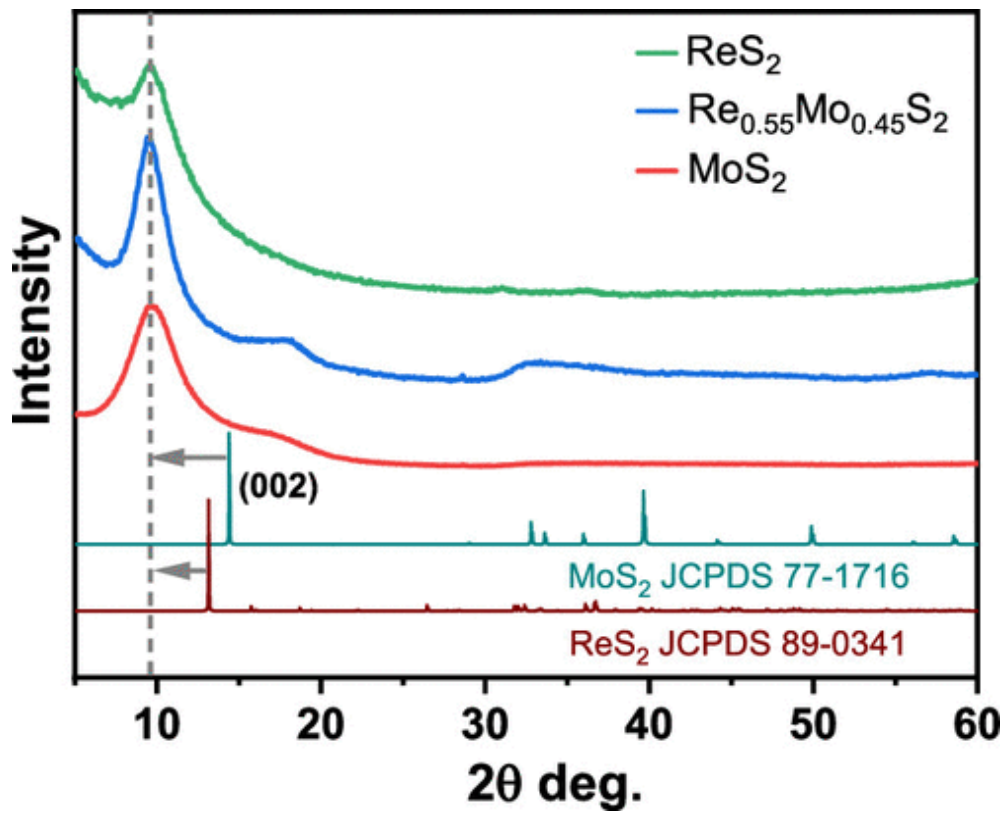

**Figure 3.** XRD patterns of the MoS$_2$, ReS$_2$, and Re$_{0.55}$Mo$_{0.45}$S$_2$ nanomaterials with enlarged interlayer distances with the corresponding bulk references of 2H MoS$_2$ (JCPDS 77-1716) and 1T′ ReS$_2$ (JCPDS 89-0341).

The composition of the synthesized materials was also analyzed by X-ray photoelectron spectroscopy (XPS). The survey XPS spectra of the MoS$_2$, ReS$_2$, and Re$_{0.55}$Mo$_{0.45}$S$_2$ nanomaterials are displayed in Figure 4a with the binding energy ranges associated with Re 4f, S 2p, and Mo 3d highlighted for clarity. Figure 4b–d displays high-resolution core-level spectra of S 2p, Mo 3d, and Re 4f measured on Re$_{0.55}$Mo$_{0.45}$S$_2$. The S 2p spectrum (Figure 4b) exhibits a single doublet at binding energies of 162.9 and 164.2 eV associated with S2p$_{3/2}$ and S2p$_{1/2}$, respectively. These binding energies are characteristic to sulfides with an oxidation state of S$^{2-}$. (11)Similarly, a single S 2s component is observed in the Mo 3d spectrum (Figure 4c), confirming that no additional sulfur species are present. A single Mo 3d doublet at 229.5 and 232.7 eV, assigned to $3d_{5/2}$ and $3d_{3/2}$, respectively, is associated with Mo in the oxidation state of 4+ . (11,37) Importantly, no higher binding energy doublet is observed, which typically arises from $MoO_3$ frequently present in $MoS_2$ samples fabricated by various methods. (38,39) The Re 4f spectrum (Figure 4d) also consists of a single doublet at binding energies of 44.7 and 42.3 eV, which correspond to $Re^{4+}$. (11,40) These results prove that in alloyed $Re_{0.55}Mo_{0.45}S_2$, all elements exist in their characteristic oxidation states with no impurity phases. Similarly, no additional species were detected in pure $ReS_2$ and $MoS_2$ nanoflakes, the high-resolution XPS spectra of which are presented in Figures S10 and S11, respectively.

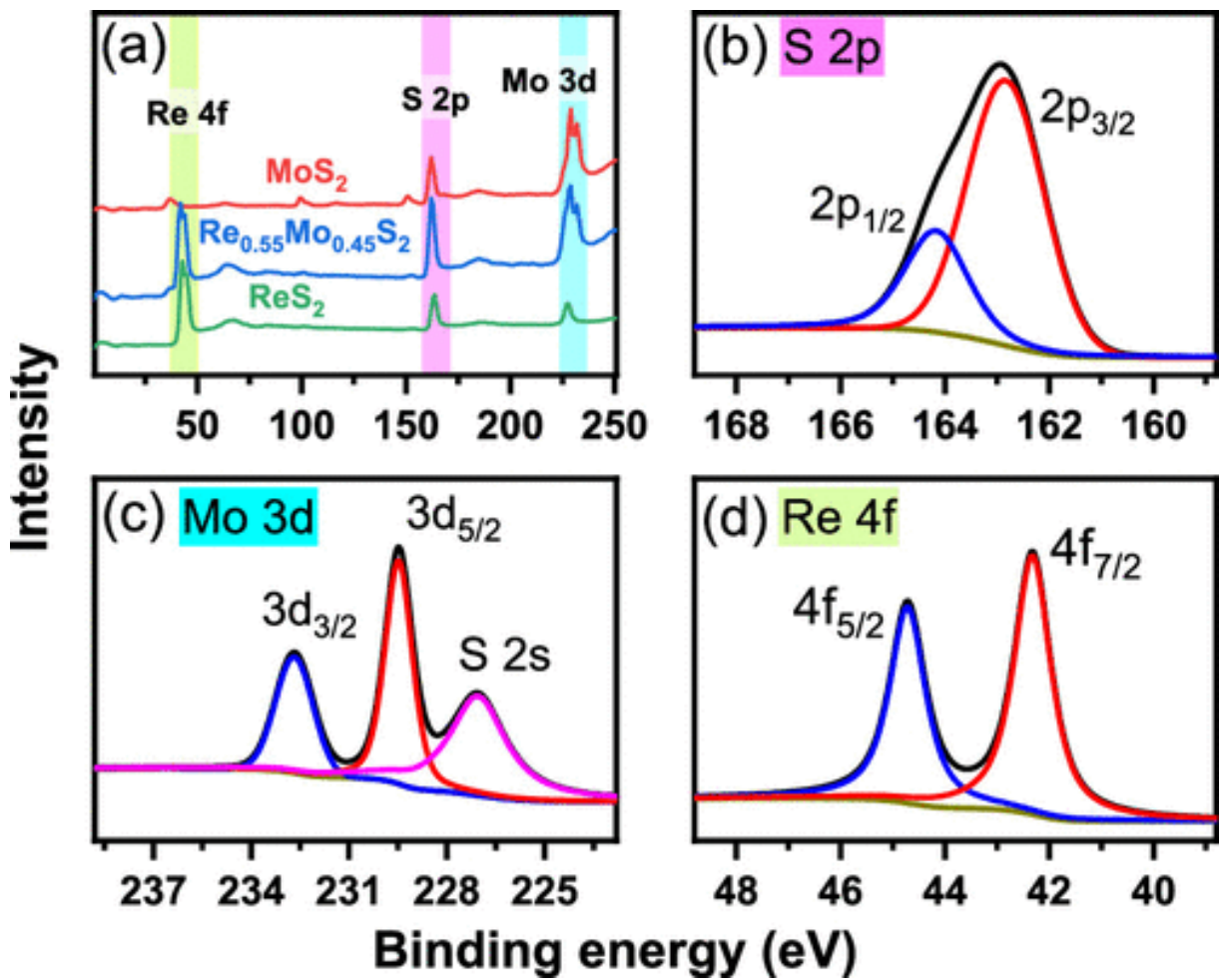

**Figure 4.** XPS survey spectra of the $ReS_2$, $MoS_2$, and $Re_{0.55}Mo_{0.45}S_2$ nanomaterials (a) and corresponding high-resolution core-level signals of S 2p (b), Mo 3d (c), and Re 4f (d) in $Re_{0.55}Mo_{0.45}S_2$ alloy nanoflakes.

**Electrochemical Characterization: HER**

The electrochemical performance of the samples coated on glassy carbon electrodes was tested in 0.5 M $H_2SO_4$ solution using a typical three-electrode measurement setup. We first characterized the catalytic performance of the pure-phase $MoS_2$, $ReS_2$, and alloyed $Re_xMo_{1-x}S_2$ nanomaterials. As seen in [Figure S12](), among different compositions of the alloys, the $Re_{0.55}Mo_{0.45}S_2$ sample exhibited the best performance. Comparing $Re_{0.55}Mo_{0.45}S_2$, $Re_{0.55}Mo_{0.45}S_2$ mixed with carbon black, and the $Re_{0.55}Mo_{0.45}S_2$ nanoflakes/carbon composite, we found that the composite exhibited the highest catalytic activity, as displayed in [Figure S13](). These results imply that the conductivity of pure $Re_{0.55}Mo_{0.45}S_2$ is the lowest among the three tested samples, while the $Re_{0.55}Mo_{0.45}S_2$ nanoflakes mixed with carbon black had worse contact compared to the $Re_{0.55}Mo_{0.45}S_2$ nanoflakes/carbon composite. [Figure 5]()a shows the polarization curves of the alloyed $Re_{0.55}Mo_{0.45}S_2$ nanoflakes/carbon composite compared with the pure-phase $MoS_2$-nanoflakes/carbon composites and $ReS_2$-nanoflakes/carbon composite recorded at a scan rate of 5 mV/s. The polarization curves exhibit remarkably different onset potentials for the three materials with the best values of 26 mV at 1 mA/cm$^2$ and 79 mV at 10 mA/cm$^2$ for $Re_{0.55}Mo_{0.45}S_2$. The Tafel slope, an important indicator for the intrinsic potential of materials in catalysis, was employed to understand the rate-limiting steps involved in HER. For this, typically, the Tafel equation of $\eta = a + b\log j$ is used, where $\eta$, $a$, $b$, and $j$ are the overpotential, Tafel constant, Tafel slope, and current density, respectively. [Figure 5]()b illustrates the Tafel slopes of $ReS_2$, $MoS_2$, and $Re_{0.55}Mo_{0.45}S_2$ nanoflakes/carbon composites in the low-potential region. The Tafel slope of the $MoS_2$/carbon composite and $Re_{0.55}Mo_{0.45}S_2$/carbon composite was calculated to be 38 mV/dec and 42 mV/dec, respectively, which is remarkably smaller than the value for the $ReS_2$/carbon composite sample. Furthermore, this Tafel slope value of the $MoS_2$/carbon composite is comparable to that of other Re-containing $MoS_2$-based HER catalysts reported in the literature (summarized in [Table S2]()), indicating the efficient kinetics of $H_2$ evolution catalyzed by the $MoS_2$ nanoflakes with enlarged interlayer distance. At the same time, the overpotential values of the $Re_{0.55}Mo_{0.45}S_2$ nanoflakes/carbon composite measured at 10 mA/cm$^2$ current density are the smallest compared to other similar Re-containing $MoS_2$-based catalysts published to date ([Figure 5]()d). (6,31,40,41) Another important characteristic of catalysts is their stability. To assess it, we performed electrochemical tests on the $Re_{0.55}Mo_{0.45}S_2$/carbon composite before and after 1000 cycles, as shown in [Figure 5]()c, which reveals that the polarization curve after cycling is very similar to the initial one, confirming excellent stability of the sample.

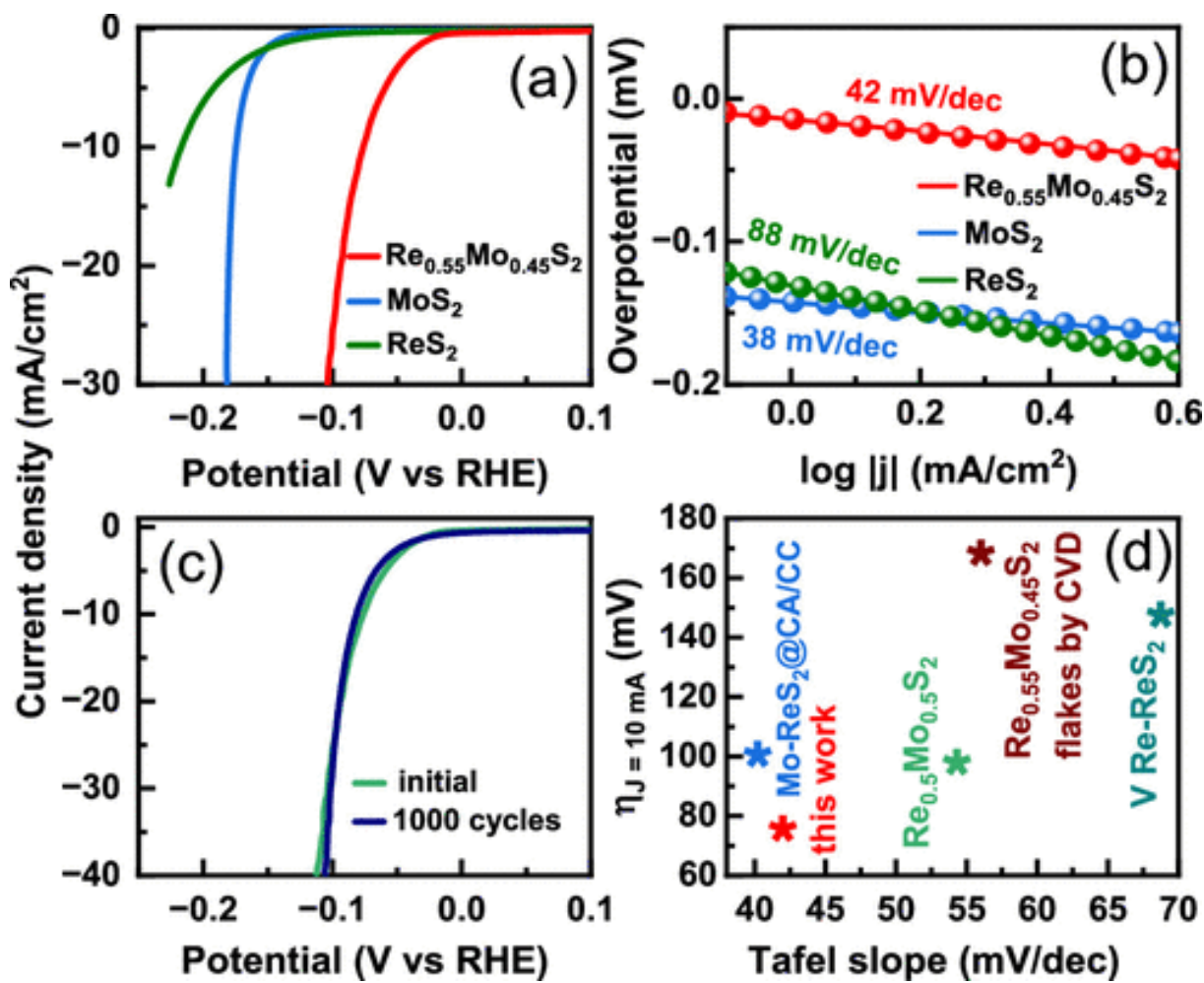

**Figure 5.** HER polarization curves of MoS$_2$, ReS$_2$, and Re$_{0.55}$Mo$_{0.45}$S$_2$ nanoflakes/carbon black composites (a) with corresponding Tafel plots (b). Stability test performed on the Re$_{0.55}$Mo$_{0.45}$S$_2$-nanoflakes/carbon black composite sample by acquiring its polarization curves before and after 1000 cycles. Comparison of overpotential at 10 mA/cm$^2$ and the Tafel slope of the Re$_{0.55}$Mo$_{0.45}$S$_2$ nanoflakes/carbon black composites with different Re-MoS$_2$-based catalysts published to date, (6,29,31,40) including Mo-ReS$_2$@CA/CC (ref (38)), Re$_{0.5}$Mo$_{0.5}$S$_2$ (ref (31)), Re$_{0.55}$Mo$_{0.45}$S$_2$ flakes by CVD (ref (29)), and V Re-ReS$_2$ (ref (6)) (d).

We finally carried out DFT calculations to understand the physical essence of the enhanced catalytic performance of the expanded Re$_{0.55}$Mo$_{0.45}$S$_2$ nanoflakes. The hydrogen adsorption free energy ($\Delta G_H$) values calculated for the expanded MoS$_2$, ReS$_2$, and Re$_{0.55}$Mo$_{0.45}$S$_2$ supercells are shown in Figure 6a. As it has been mentioned in the introduction, an efficient catalyst for HER should possess a $\Delta G_H$ as close as possible to 0 eV. (42) As follows from the results, the $\Delta G_H$ value of the expanded Re$_{0.55}$Mo$_{0.45}$S$_2$ is much closer to the ideal value of 0 eV, as compared to that of the expanded MoS$_2$ or expanded ReS$_2$ supercells, which proves that the expanded alloy nanoflakes should possess the strongest HER activity. The distribution of the density of states (DOS) of the normal and expanded $Re_{0.55}Mo_{0.45}S_2$ supercells presented in Figure 6b indicates that the expanded supercell has a slightly larger number of DOS than the normal one, which means that expanded $Re_{0.55}Mo_{0.45}S_2$ possesses higher electronic transmission capability. (17) Moreover, it is known that the catalytic activity is closely related to the positions of the d-band center with respect to the Fermi level. (43) The DOS of the expanded $MoS_2$, $ReS_2$, and $Re_{0.55}Mo_{0.45}S_2$ supercells are shifted upward compared to that of the corresponding normal supercells with an obvious shift of the d-band centers toward the Fermi level (Figure S14). This suggests that expanded $MoS_2$, $ReS_2$, and $Re_{0.55}Mo_{0.45}S_2$ nanoflakes more easily provide electrons in HER as compared to the corresponding normal compounds, leading to their stronger catalytic activity.

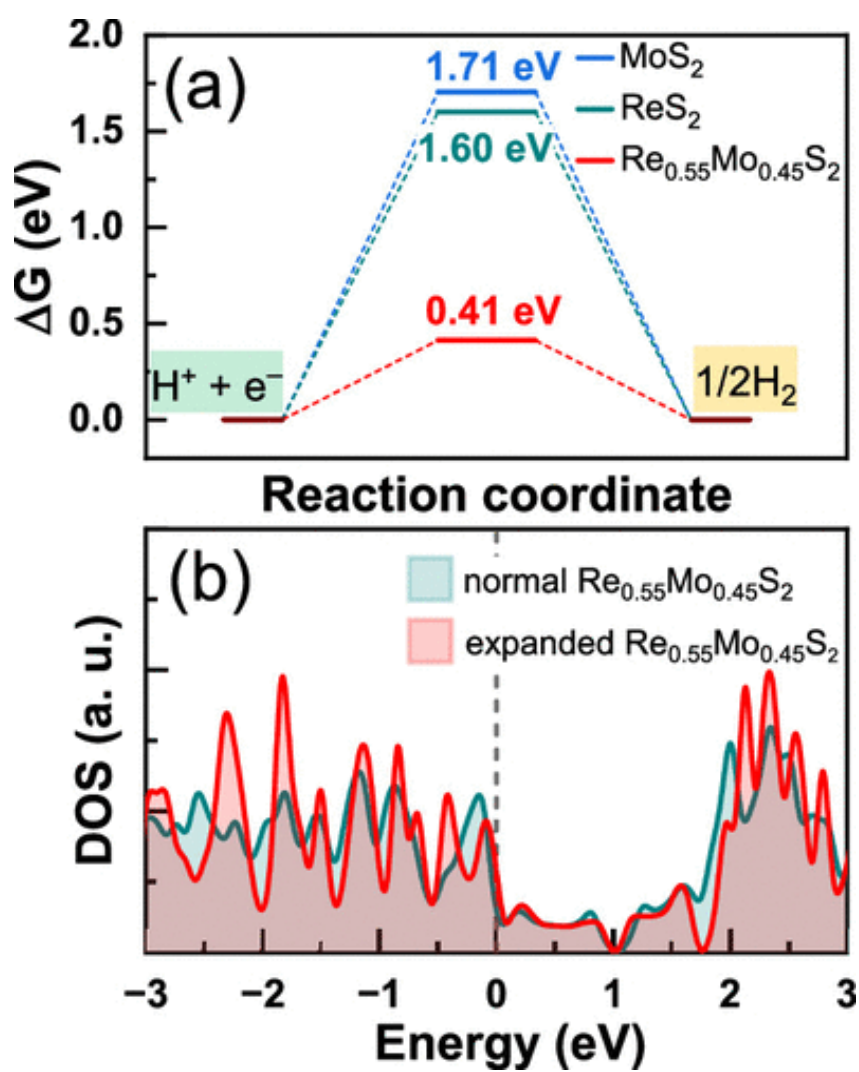

**Figure 6.** Hydrogen adsorption free energies of the expanded $MoS_2$, $ReS_2$, and $Re_{0.55}Mo_{0.45}S_2$ supercells (a). DOSs for the normal and expanded $Re_{0.55}Mo_{0.45}S_2$ supercells (b).

**Conclusions**

We developed a facile colloidal synthesis of pure-phase MoS$_2$, ReS$_2$, and alloyed Re$_x$Mo$_{1-x}$S$_2$ nanomaterials with enlarged interlayer spacing, employing a simple deaeration procedure that does not require degassing in vacuum. Polar formamide used as a solvent can form donor–acceptor complexes with metal precursors, which contribute to the enlarged interlayer distance in growing nanoflakes. At the same time, its presence on the surface does not block active sites of the nanomaterials available for catalysis. In addition, the size of the nanoflakes can be controlled by employing different sulfur precursors, where nanoflakes synthesized with thiourea exhibit the largest lateral dimensions, while samples obtained with alkali metal sulfides had smaller sizes. Re$_{0.55}$Mo$_{0.45}$S$_2$ nanoflakes with enlarged interlayer distances demonstrated excellent electrochemical performance characterized by a small overpotential of 79 mV at 10 mA/cm$^2$ and a small Tafel slope of 42 mV/dec. DFT calculations proved that in such an alloyed compound with enlarged interlayer distances, the Gibbs free energy of the hydrogen adsorption can significantly be lowered. In this way, its electronic structure can be modulated, resulting in a remarkable enhancement of its catalytic performance in HER. We believe that the developed synthesis strategy can pave the way to efficient and inexpensive TMD-based catalysts for hydrogen generation.

**Experimental Section**

**Chemicals**

Sodium sulfide (Na$_2$S, ≤100%), lithium sulfide (Li$_2$S, 99.95%), thiourea (>99.0%), molybdenum chloride (MoCl$_5$, 95%), rhenium chloride (ReCl$_5$, ≤100%), formamide (FA, 99.5%), dimethylformamide (DMF, 99.8%), and *N*-methylformamide (MFA, 99%) were purchased from Acros Organics; potassium sulfide (K$_2$S, 99%) was purchased from American Elements; carbon black (Vulcan XC-72R) was purchased from Fuel Cell Earth; isopropanol (99%) was purchased from VWR; ethanol (anhydrous, >99%) was purchased from Merck; chloroform (99.9%) was purchased from ROTH. For the ICP-OES measurements, rhenium AAS standard solution (Re 1000 μg/mL) and molybdenum plasma standard solution (Mo 1000 μg/mL) were purchased from Alfa Aesar.

*Synthesis of the MoS$_2$ and ReS$_2$ Nanomaterials*

To synthesize the MoS$_2$ (or ReS$_2$) nanoflakes, 20 mg (0.07 mmol) of MoCl$_5$ (or 25.5 mg (0.07 mmol) of ReCl$_5$, respectively) were dissolved in 3 mL of FA, whereupon the color of the solution quickly turned orange, followed by bubbling with nitrogen for 30 min to remove air.

In parallel, 40 mg (0.5 mmol) of Na$_2$S (or an equivalent amount of another S-precursor) were dissolved in 30 mL of FA in a three-neck 100 mL flask upon bubbling with nitrogen for 30 min. Then, the flask with the S-precursor was heated up to 170 °C under a nitrogen atmosphere, and the Mo-precursor solution was added from a syringe pump at a rate of 3 mL/min upon vigorous stirring. The color of this mixture quickly changed from orange to black upon adding the Mo-precursor. After 1 h at 170 °C, the heating mantle was removed and the reaction mixture was cooled down to room temperature. The resulting MoS$_2$ (respectively ReS$_2$) nanoflakes were purified by adding 40 mL of isopropanol to the reaction mixture followed by centrifugation at 5200 rpm for 5 min. The precipitate was redispersed in 10 mL of ethanol, precipitated by centrifugation two times, and kept in 10 mL of ethanol.

*Synthesis of the Re$_x$Mo$_{1-x}$S$_2$ Alloy Nanoflakes*

To synthesize the alloyed Re$_x$Mo$_{1-x}$S$_2$ nanoflakes, Mo- and Re-precursor solutions with different ratios between MoCl$_5$ and ReCl$_5$ (Mo/(Mo + Re) = 0, 0.2, 0.4, 0.5, 0.6, 0.8, and 1) were prepared by dissolving the salts in 3 mL of FA followed by deaeration with nitrogen bubbling for 30 min. The next steps were the same as in the synthesis of the pure phase nanomaterials.

*Synthesis of the TMD Nanoflakes/Carbon Composites*

To synthesize the composite, first, 3 mg of carbon black were dispersed in 30 mL of FA upon sonication for 5 min to obtain a homogeneous mixture. Then, 40 mg of Na$_2$S were dissolved in 30 mL of FA in a three-neck 100 mL flask upon bubbling with nitrogen for 30 min. In parallel, 10 mg of MoCl$_5$ and 15.5 mg of ReCl$_5$ were dissolved in 3 mL of FA upon bubbling with nitrogen for 30 min. Then, the flask with the S-precursor and carbon black was heated up to 170 °C under a nitrogen atmosphere, and the Mo-/Re-precursor was added from a syringe pump at a rate of 3 mL/min upon vigorous stirring. After 1 h at 170 °C, the Re$_{0.55}$Mo$_{0.45}$S$_2$-nanoflakes/carbon composite was purified by adding 90 mL of ethanol and 30 mL of chloroform to the reaction mixture followed by centrifugation at 12000 rpm for 10 min. The precipitate was purified two times by redispersing it in 30 mL of ethanol, followed by ultrasonication and centrifugation. The product was dispersed in 10 mL of ethanol. To synthesize the MoS$_2$ and ReS$_2$/carbon composites, 20 mg of MoCl$_5$ were used for MoS$_2$ and 25.5 mg of ReCl$_5$ were used for ReS$_2$. The next steps were the same as in the synthesis of alloyed Re$_{0.55}$Mo$_{0.45}$S$_2$ nanoflakes/carbon composite.

**Characterization**

Transmission Electron Microscopy

Bright-field TEM imaging on a Zeiss Libra 120 transmission electron microscope operated at 120 kV was used to characterize the morphology of the nanomaterials. High-angle annular dark-field scanning TEM (HAADF-STEM) imaging and spectrum imaging analysis based on energy-dispersive X-ray spectroscopy (EDXS) were performed on a Talos F200X microscope equipped with an X-FEG electron source and a Super-X EDX detector system (FEI) at 200 kV. For TEM specimen preparation, the corresponding nanoflakes were purified three times with ethanol and then drop-cast from diluted ethanol dispersions onto carbon-coated copper grids.

Scanning Electron Microscopy

SEM imaging was used to analyze the morphology and roughness of the films made of pure $Re_{0.55}Mo_{0.45}S_2$ nanoflakes, $Re_{0.55}Mo_{0.45}S_2$ nanoflakes mixed with carbon black, and the $Re_{0.55}Mo_{0.45}S_2$ nanoflakes/carbon composite. It was performed on a ZEISS Gemini SEM 300 Microscope. The specimens were prepared by drop-casting an ink, which also was used for electrochemical characterization, onto a silicon wafer.

*Raman Spectroscopy*

Raman spectra of the nanomaterials were acquired on a high-resolution confocal Raman microscope MonoVista CRD+ (S&I Spectroscopy & Imaging GmbH) with a liquid nitrogen-cooled CCD PyLoN: 100 BRX detector (Princeton Instruments). The samples were prepared by drop-casting nanoflakes from ethanol on a microscope slide followed by drying in air. The spectra were recorded with 514 nm laser excitation and a power of approximately 0.6 mW on the sample.

*X-ray Diffraction Analysis*

Powder X-ray diffraction patterns were collected in reflection mode with a Bruker AXS D2 PHASER instrument equipped with a nickel filter and a LYNXEYE/SSD160 detector using Cu K$\alpha_{1,2}$ irradiation.

*Inductively Coupled Plasma Optical Emission Spectroscopy Analysis*

ICP-OES was employed to analyze the composition of the nanomaterials on a PerkinElmer Optima 7000DV system. To prepare the samples, the nanoflakes were precipitated from their 10 mL dispersions in ethanol, additionally purified with ethanol, and then dried under a nitrogen atmosphere for 2 h. The powders were decomposed by adding first 500 μL of HF (40% concentration) and then, after 10 min, adding 1000 μL of HNO$_3$ (60% concentration) and keeping the mixture overnight. Thereafter, the solution obtained was diluted to 20 mL volume with ultrapure water.

*X-ray Photoelectron Spectroscopy*

The samples for XPS analysis were prepared via drop-casting the nanoflakes on a silicon wafer. Then, the samples were transferred to an ultrahigh vacuum chamber (ESCALAB 250Xi by Thermo Scientific, base pressure: $2 \times 10^{-10}$ mbar) for the XPS analysis. The measurements were carried out using an XR6 monochromatized Al K$_\alpha$ source ($hv$ = 1486.6 eV) and a pass energy of 20 eV. The binding energy scale was internally referenced to the C 1s peak at 285 eV.

*Electrochemical Tests*

A three-electrode measurement setup was used to carry out the electrochemical tests. The setup consists of a reference electrode (Ag/AgCl in saturated KCl), a counter electrode (Pt), and a glassy carbon disk working electrode with a diameter of 3 mm (geometric area of 0.07 cm$^2$). A sample kept in 10 mL ethanol was first sonicated for homogenization, and then 1.5 mL of this dispersion was centrifuged. The precipitate was dispersed in 400 µL of an isopropanol/water mixture (1/3 vol) and sonicated for 1 h followed by stirring for 1 h. Subsequently, the thus-obtained dispersion was drop-cast on the surface of the glassy carbon electrode and dried in air. This deposition was repeated three times to form a thicker layer of the nanomaterials, amounting for about 0.247 mg/cm$^2$.

*Computational Details*

All the first-principle calculations based on DFT were carried out by the VASP.5.4 (44) program. The projector augmented wave (PAW) (45) method was used to treat the core–valence electron interaction. The generalized gradient approximation (GGA) (46) in the form of the Perdew–Burke–Ernzerhof (PBE) (47) functional was adopted to describe the exchange-correlation potential. The DFT-D3 method proposed by Grimme et al. (26) was utilized to describe the weak van der Waals (vdW) interactions. The cutoff energy, the energy convergence criterion, and the force convergence criterion were set as 500 eV, $10^{-6}$ eV, and $10^{-2}$ eV/Å, respectively. In addition, a $2 \times 2$ supercell of normal interlayer distance MoS$_2$ and a $1 \times 1$ supercell of normal ReS$_2$ were employed. Considering the complexity of the calculations, the normal Re$_{0.55}$Mo$_{0.45}$S$_2$ model was simplified to normal Re$_{0.5}$Mo$_{0.5}$S$_2$ constructed by replacing half of the Re atoms by Mo atoms in a $1 \times 1$ supercell of normal ReS$_2$. The expanded MoS$_2$, ReS$_2$, and Re$_{0.55}$Mo$_{0.45}$S$_2$ models were built by enlarging the $d$ spacing distance to 9.22 Å based on the normal MoS$_2$, ReS$_2$, and Re$_{0.55}$Mo$_{0.45}$S$_2$ models, respectively. The $6 \times 6 \times 3$, $6 \times 6 \times 6$, and $6 \times 6 \times 6$ Monkhorst–Pack $k$-points were respectively utilized to integrate the Brillouin zones for normal $MoS_2$, $ReS_2$, and $Re_{0.55}Mo_{0.45}S_2$, while the $6\times6\times2$, $6\times6\times4$, and $6\times6\times4$ Monkhorst–Pack $k$-points were respectively used to integrate the Brillouin zones for expanded $MoS_2$, $ReS_2$, and $Re_{0.55}Mo_{0.45}S_2$. The related structure models and the structure models with an H atom absorbed on the materials are shown in Figures S15 and S16.

The free energy can be computed by the method developed by Norskov et al: (48,49)

$$\Delta G = \Delta E + \Delta E_{ZPE} - T\Delta S \quad (1)$$

where $\Delta E$, $\Delta E_{ZPE}$, $\Delta S$, and $T$ denote the total energy change, the zero point energy change, the entropy change, and the system temperature, respectively. Then, the atomic hydrogen adsorption free energy ($\Delta G_H$) was computed by:

$$\Delta G_H = G_{*H} - 0.5 G_{H2} - G_* \quad (2)$$

where $G_{*H}$, $G_{H2}$, and $G_*$ denote the free energy of a hydrogen atom on the adsorption site, the free energy of hydrogen gas, and the free energy of the adsorption site, respectively.

## Acknowledgements

We thank Dr. A. Antanovich (TU Dresden) for assistance in ICP-OES measurements, S. Goldberg (TU Dresden) for assistance in SEM imaging, and Prof. N. Gaponik (TU Dresden) for fruitful discussions. J.L. acknowledges the China Scholarship Council (no. 202006750019). A.B. acknowledges the Graduate Academy of TU Dresden for providing the "Scholarship for the promotion of early career female scientists of TU Dresden". The use of the HZDR Ion Beam Center TEM facilities and the funding of TEM Talos by the German Federal Ministry of Education and Research (BMBF), grant no. 03SF0451, in the framework of HEMCP are acknowledged. Y.V. has received funding from the European Research Council (ERC) under the European Union's Horizon 2020 research and innovation program (ERC grant agreement no. 714067, ENERGYMAPS).## References